# Modelling human seat contact interaction for vibration comfort

**Raj Desai, Marko Cvetković, Georgios Papaioannou, Riender Happee**

Cognitive Robotics (CoR) Delft University of Technology Netherlands

## ABSTRACT

The seat to head vibration transmissibility depends on various characteristics of the seat and the human body. One of these, is the contact interaction, which transmits vibrational energy from the seat to the body. To enhance ride comfort, seat designers should be able to accurately simulate seat contact without the need for extensive experiments. Here, the contact area, pressure, friction and seat and body deformation in compression and shear play a significant role. To address these challenges, the aim of this paper is to define appropriate contact models to improve the prediction capabilities of a seated human body model with regards to experimental data. A computationally efficient multibody (MB) model is evaluated interacting with finite element (FE) and MB backrest models, using several contact models. Outcomes are evaluated in the frequency domain for 3D vibration transmission from seat to pelvis, trunk, head and knees. Results illustrate that both FE and MB backrest models allowing compression and shear provide realistic results.

## Introduction

Since occupants' well-being is directly affected by comfort, it is crucial to design automotive seating with the consideration of comfort. By simulating human body-seat interactions, designers can gather data and insights that inform the selection of materials, dimensions, ergonomics, and other design elements. These interactions can be modelled using multibody and/or finite element (FE) models. Finite element analysis (FEA) can provide detailed insights into the interactions between the human body and seating, including deformation in compression and shear. However, FEA can be computationally intensive and time-consuming, especially when simulating complex models with numerous elements and conducting multiple iterations (Roupa et al., 2022).

Contact interaction, which involves the transmission of vibrations from the seat to the human body, is a critical factor in determining ride comfort. Accurate modelling of contact interactions (Desai et al., 2021) is challenging due to the multiple characteristics involved, such as contact area, seat or body soft tissue deformations, dynamic friction, damping, pressure, and shear stress (Mirakhorlo, Kluft, Shyrokau, et al., 2022). The vertical forces are generated due to the weight of the occupant and the cushion (Papaioannou et al., 2020), while the shear forces, which can also cause discomfort, are affected by the interaction between the human body and the seat cushion. Therefore, to improve ride comfort, seats that effectively isolate shear forces to the occupant are needed. The aim of this paper is to explore appropriate contact models (capturing shear) that can (a) capture the dependency on various characteristics of the seat and the human body, (b) improve the prediction of body motion (seat-to-head vibrational transmissibility) in an erect seated position based on experimental vibration data (Mirakhorlo, Kluft, Shyrokau, et al., 2022).

## Method

A computationally efficient human body model (EHM) (Desai et al., 2023) consisting of 12 segments is presented in Fig. 1. The seat-human body integrated system is modelled in MADYMO (Tass, 2019) and validated using 3D human response data (Mirakhorlo, Kluft, Shyrokau, et al., 2022). The (force-displacement/stress-strain comression) contact model is investigated using various combinations of multibody surfaces (ellipsoids), rigid 3D surfaces (facets) and deformable finite elements (FE). Three backrest models are investigated representing the two foam blocks of the experimental seat. An FE backrest was made up of 1523 four node tetra elements. The foam characteristics were defined using experimental load/unloading functions, hysteresis slope and density. Contact was established with a friction coefficient of 1.2. Two multibody (MB) backrest models are presented. The $MB_{friction}$ model applies standard friction which provides poor results in lateral loading in particular. The model drifted in lateral loading and this was resolved by adding weak springs preventing lateral drift at thighs and pelvis. However the friction still prevented a realistic shear deformation of seat and human which was well captured by the FE backrest. This was resolved in the model $MB_{shear}$, where contact friction was removed and replaced by 2D point restraints acting in the contact plane at thighs, pelvis and back with linear stiffness and damping. The human model postural stabilization parameters for the three backrest models were obtained using optimization, where for model $MB_{shear}$, also the point restraint stiffness and damping were tuned.

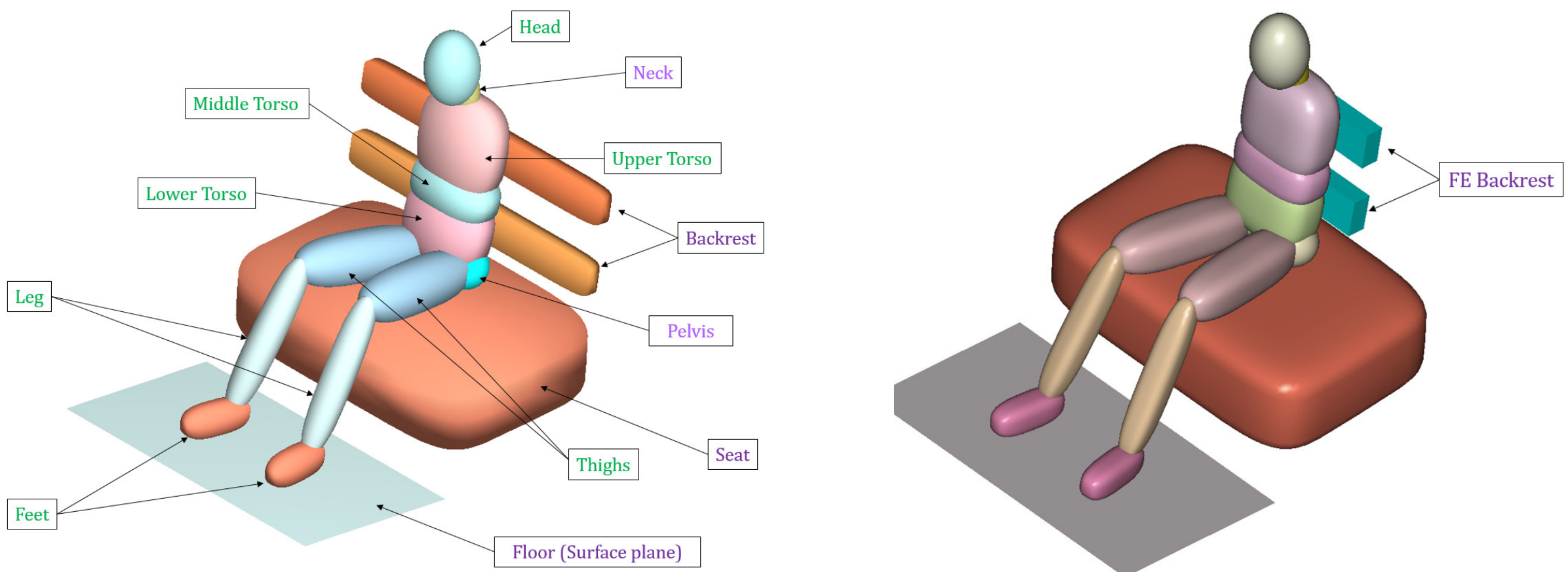

Fig. 1 Human body model (left: MB backrest, right: FE backrest)

## Results

The experimental and model results are presented in Fig. 2 - Fig. 4. In comparison to the FE backrest, the MB backrest with shear can achieve comparable outcomes with less computing effort. The computational time required for simulating 35 seconds perturbation (with time step of 1E-3s) was 3.5 times more in the EHM with FE backrest compared to the EHM with MB backrest, which runs almost in real time. During vertical excitations, such as vibrations or impacts acting vertically on the seat backrest, the deformations in the FE elements can lead to the generation of higher forces. As a result, the FE elements may experience larger deformations, leading to higher forces being transmitted from the backrest to the torso, and consequently resulting in higher seat-to-head transmissibility (Fig. 4). Lower angular rotations are present in fore-aft and lateral excitations with the $MB_{friction}$ model which allows insufficient trunk motion (Fig. 2-3). The results of $MB_{shear}$ and $MB_{friction}$ in the vertical loading case are comparable, showing a small influence of shear.

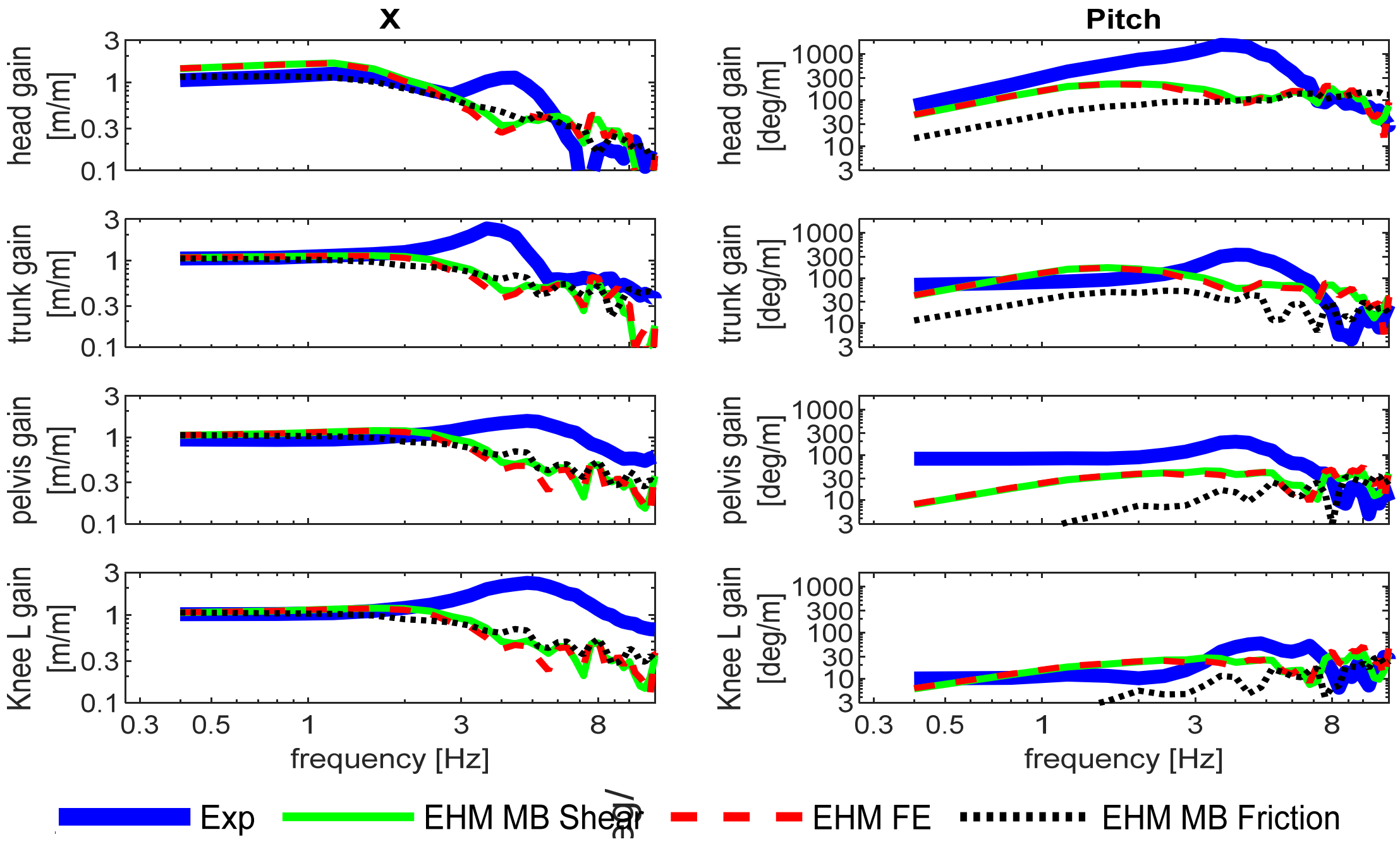

**Fig. 2:** Model simulation results in fore-aft loading (EHM MB: EHM with Multibody backrest, EHM FE: EHM with Finite Element backrest).

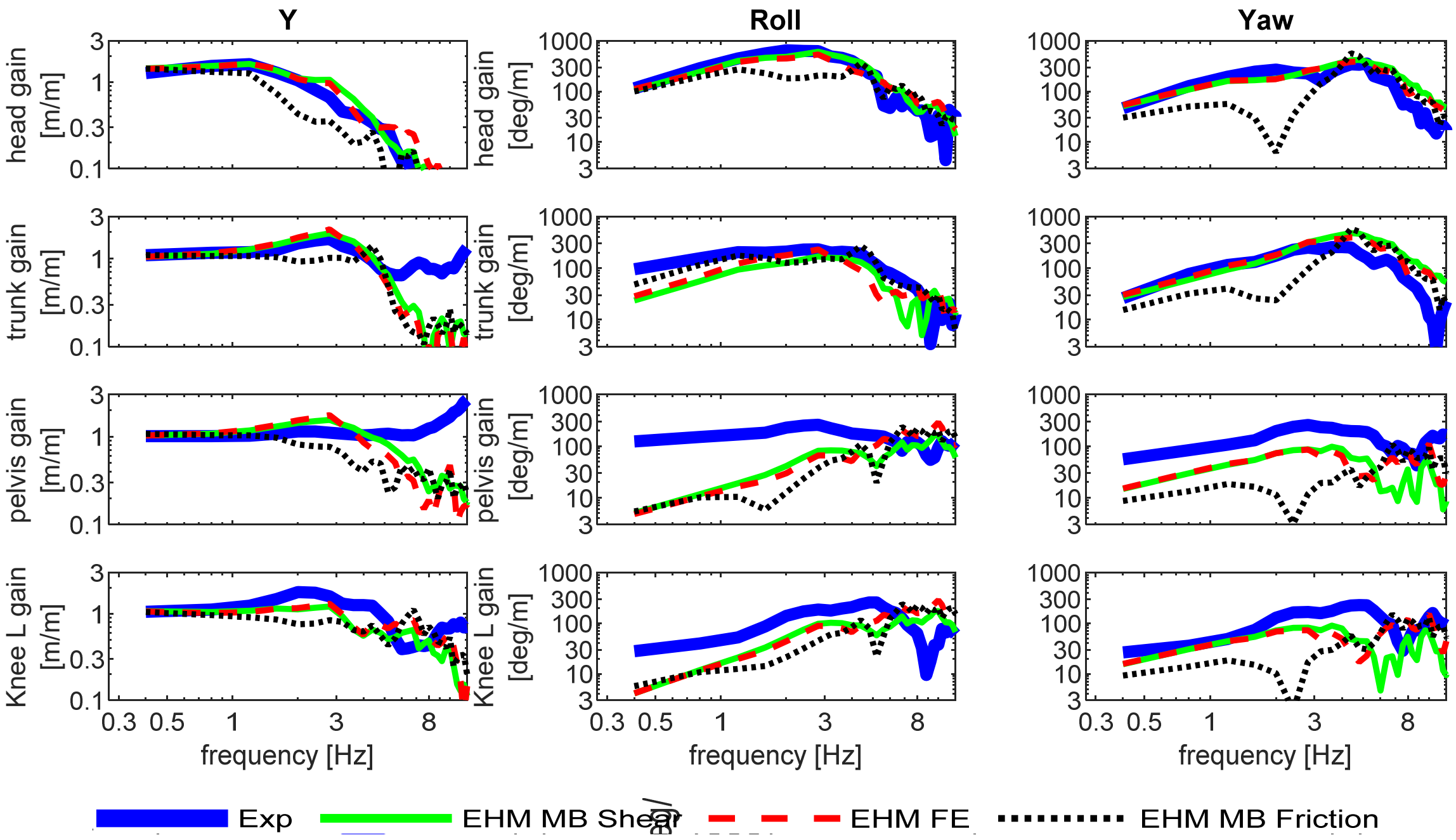

**Fig. 3:** Model simulation results in lateral loading.

## Conclusion

The results show that the vibration transmissibility to the occupants is significantly influenced by the contact model. This paper investigates the characterization of the physical interactions between the human body and the seat backrest during motion precisely. The force-type contact model is the most efficient in enhancing the model's body motion prediction with respect to experimental data. Although using FE backrest increases model accuracy, it also increases computing load. The MB model with point restraint contact model allowing shear captures the interactions efficiently and accurately. By studying the interaction between the human body and the seat backrest, the paper provides valuable insights into how to efficiently capture the shear interactions. The findings of this study may also be used to guide the design of seats for improved comfort. In the future, we intend to precisely model interactions while taking into account the geometry/shape of the seat, buttocks, and backrest using pressure data.

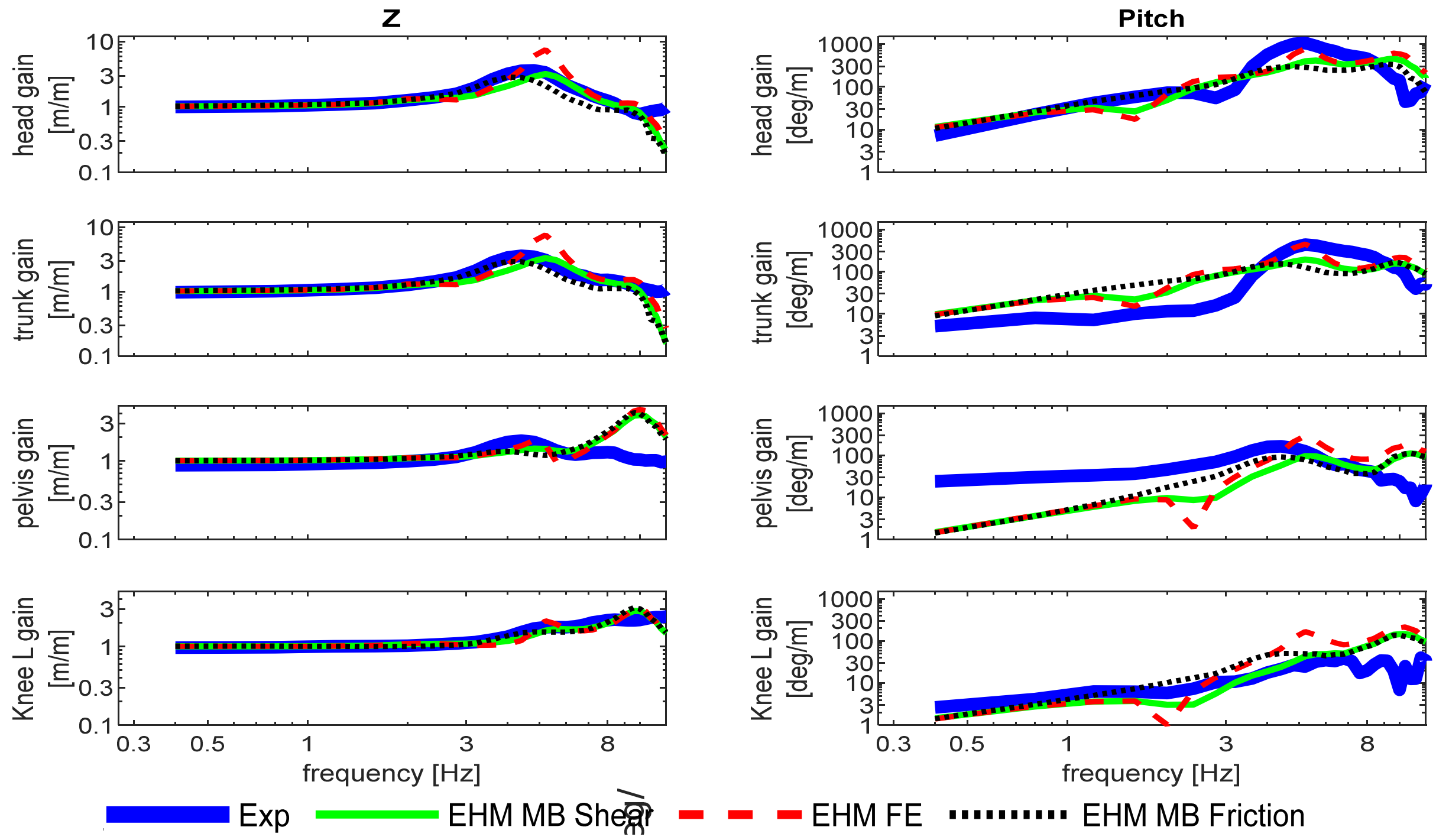

**Fig. 4:** Model simulation results vertical loading case.

**Acknowledgement**: We acknowledge the support of Toyota Motor Corporation.